\documentclass
[%
aps,
reprint,
showpacs,
preprintnumbers,
amsmath,
amssymb,
pre,
]{revtex4-1}

\usepackage{graphicx}% Include figure files
\usepackage{dcolumn}% Align table columns on decimal point
\usepackage{bm}% bold math
\usepackage{subfigure}
\usepackage{float}
\usepackage{color,soul}
\usepackage{stfloats}
\begin{document}
	%\preprint{APS/123-QED}
		
	\setulcolor{red}    %set underlining color
	\setstcolor{blue}   %set overstriking color
	\sethlcolor{yellow} %set highlighting color
	%\captionsetup[figure]{labelfont={bf},name={Fig.},labelsep=period}
	\title{Polarized proton acceleration in ultra-intense laser interaction with near critical density plasmas}
	
	\author{X. F. Li$^{1,2,3}$}
	\author{P. Gibbon$^{1,4}$}\email{p.gibbon@fz-juelich.de}%
	\author{A. H\"{u}tzen$^{5,6}$}%
	\author{M. B\"{u}scher$^{5,6}$}
	\author{S. M. Weng$^{2,3}$}\email{wengsuming@gmail.com}%
	\author{M. Chen$^{2,3}$}
	\author{Z. M. Sheng$^{2,3,7,8}$}
	\affiliation{$^1$ Institute for Advanced Simulation,J\"{u}lich Supercomputing Centre, Forschungszentrum J\"{u}lich, 52425 J\"{u}lich,Germany}
	\affiliation{$^2$ Key Laboratory for Laser Plasmas (MoE), School of Physics and Astronomy, Shanghai Jiao Tong University, Shanghai 200240, China}%
	\affiliation{$^3$ Collaborative Innovation Center of IFSA, Shanghai Jiao Tong University, Shanghai 200240, China}%
    \affiliation{$^4$ Centre for Mathematical Plasma Astrophysics,
	Katholieke Universiteit Leuven, 3000 Leuven, Belgium}
	\affiliation{$^5$ Peter Gr\"{u}nberg Institut (PGI-6), Forschungszentrum J\"{u}lich, 52425 J\"{u}lich, Germany }
	\affiliation{$^6$Institut f\"{u}r Laser- und Plasmaphysik, Heinrich-Heine-Universit\"{a}t D\"{u}sseldorf, 40225 D\"{u}sseldorf, Germany}
	\affiliation{$^7$ SUPA, Department of Physics, University of Strathclyde, Glasgow G4 0NG, UK}
	\affiliation{$^8$ Tsung-Dao Lee Institute, Shanghai Jiao Tong University, Shanghai 200240, China}

	\date{\today}% It is always \today, today,
	
	\begin{abstract}
		The production of polarized proton beams with multi-GeV  energies in ultra-intense laser interaction with  targets is studied with three-dimensional Particle-In-Cell simulations. A near-critical density plasma target with pre-polarized proton and tritium ions is considered for the proton acceleration. The pre-polarized protons are initially accelerated by laser radiation pressure before injection and further acceleration in a bubble-like wakefield. The temporal dynamics of proton polarization is tracked via the T-BMT equation, and it is found that the proton polarization state can be altered both by the laser field  and the magnetic component of the wakefield. The dependence of the proton acceleration and polarization on the ratio of the ion species is determined, and it is found that the protons can be efficiently accelerated as long as their relative fraction is less than $20\%$, in which case the bubble size is large enough for the protons to obtain sufficient energy to overcome the bubble injection threshold.  
	\end{abstract}
	
	%\pacs{52.38.Kd, 41.75.Jv, 52.59.Fn, 52.65.Rr}
	%52.38.Kd   Laser-plasma acceleration of electrons and ions (see also 41.75.Jv Laser-driven acceleration in electromagnetism; electron and ion optics)
	%41.75.Jv   Laser-driven acceleration (see also 52.38.-r Laser-plasma interactions in plasma physics)
	%52.50.Jm   Plasma production and heating by laser beams (laser-foil, laser-cluster, etc.)
	%52.65.Rr   Particle-in-cell method
	
	%52.59.Fn 	Multistage accelerated heavy-ion beams
	
	% PACS, the Physics and Astronomy Classification Scheme.
	
	%\keywords{Bubble regime, proton acceleration and polarization, PIC simulation and T-BMT equation.}%Use showkeys class option if keywordp
	\maketitle
	
\section{Introduction}
    With the development of laser technology, especially chirped pulse amplification (CPA)\cite{strickland1985compression}, remarkable progress has been achieved in the field of laser-plasma acceleration \cite{esarey2009physics,macchi2013ion}. Since laser-driven wakefield acceleration (LWFA) was first proposed by Tajima and Dawson in 1979 \cite{tajima1979laser}, electron beams with quasi-monoenergetic peaks  up to 7.8 GeV have been generated using a peak laser power of 850 TW interacting with a 20 cm capillary charge waveguide in 2019 \cite{gonsalves2019petawatt}. The maximum energy for laser-driven ion acceleration is around 85 MeV, via a high energy laser incident on micrometer thick plastic targets\cite{wagner2016maximum}. Near-100 MeV protons were obtained through a hybrid scheme of radiation pressure and sheath acceleration \cite{higginson2018near}. These plasma-based accelerators have prompted a new class of diagnostic techniques  different from those common to traditional accelerators\cite{downer2018diagnostics}. In order to effectively utilize and develop laser plasma acceleration, many characteristics of particle beams, \emph{e.g.} energy spread, charge, pulse duration and emittance, have been steadily improved. However, the polarization of particles has only rarely been studied in plasma-based acceleration\cite{buscher2020generation}.

    Polarization is defined as the collective spin state of a particle beam. It is commonly employed in nuclear physics\cite{anthony1993determination}, high energy physics\cite{olsen1959photon} and material physics\cite{schultz1988interaction}. For laser-driven accelerators, there are two approaches  leading to polarized particle beams: a \textit{de novo} polarization build-up from an unpolarized target in the interaction with laser, or polarization preservation of pre-aligned spins during the acceleration. In the first instance, the spin polarization of an ultra-relativistic electron beam has been investigated through colliding with an intense laser pulse in the quantum radiation-dominated regime\cite{li2019ultrarelativistic}. A possible way for producing highly polarized positron beams was also proposed via interaction of an ultra-relativistic electron beam with counterpropagating two-color intense laser pulses\cite{chen2019polarized}. For a pre-polarized target, it has been recently demonstrated that nuclear and electron spin-polarized H and D densities of at least $10^{19}cm^{-3}$ with 10ns lifetimes can be produced by photo-dissociation of HBr and DI with circularly polarized UV light pulses \cite{sofikitis2018ultrahigh}. Using this kind of pre-polarized target,  M. Wen \emph{et al} demonstrated that kiloampere polarized electron beams can be produced through LWFA in the bubble regime\cite{wen2019polarized}. Furthermore, a vortex laser interacting with pre-polarized plasma was proposed to produce energetic electrons with high polarization and beam charge\cite{wu2019polarized}.

    Because the proton is 1836 times heavier than the electron or positron, it is much more difficult to align the spin of protons via laser-solid interaction, as demonstrated at the ARCturus laser facility in D\"{u}sseldorf\cite{raab2014polarization,huetzen2019polarized,buscher2019polarized}. For this reason, acceleration of polarized proton beams has been investigated using a pre-polarized target composed of two different ion species close to the critical density for a Ti:Sapphire laser. By irradiating pre-polarized monatomic gases from photo-dissociated hydrogen halide molecules with a petawatt laser, proton beams with nearly 100 MeV energy and $80\%$ polarization  via magnetic vortex acceleration (MVA) mechanism \cite{PhysRevE.102.011201} were predicted. In order to obtain several GeV protons from near-critical plasma targets, wakefield acceleration offers a good option. The work of B. F. Shen \emph{et al} showed that the protons can be trapped and accelerated efficiently in the bubble regime, where the plasma consists a small proportion of protons and a larger proportion of heavier tritium ions\cite{shen2007bubble}. It should be noted that the density ratio of protons and heavier ions is different in the above two studies. Besides, the protons need to  be pre-accelerated to a sufficient energy in order to be injected into the wakefield. In the research of M. Liu \emph{et al}, with pre-accelerated by the radiation pressure acceleration (RPA) in a thin solid foil, the protons were captured by the LWFA in an underdense gas\cite{liu2018efficient}. 
    
    In this paper, pre-acceleration of protons via direct laser acceleration (DLA) with an ultra-intense, circularly polarized laser is proposed. This initial phase is followed by proton acceleration in a wakefield, where the polarization dynamics are studied in detail. Finally, the influence of the proton:ion ratio on the acceleration and polarization is investigated. 
    
     \begin{figure}[h]
     	\includegraphics[width=0.48\textwidth]{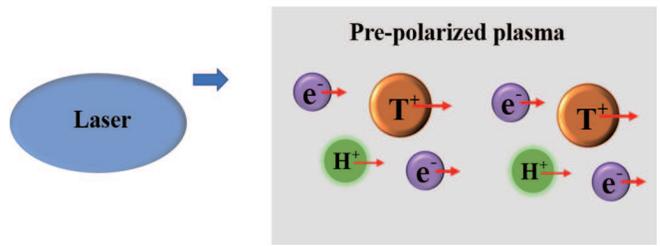}
     	\centering
     	\caption{
     		Schematic of laser pre-polarized plasma interaction. The polarization direction of particles are denoted as red arrows, which is aligned with the propagation direction of laser. The plasma is composed of electrons($e^-$), protons($H^+$) and tritium ions ($T^+$).
     	}\label{Figure1}
     \end{figure}	
	
\section{Simulation method }
	A series of simulations were performed to study the polarization dynamics during the acceleration of pre-polarized protons  with a modified version of the 3D particle-in-cell (PIC) code EPOCH \cite{arber2015contemporary}. A laser pulse with $\lambda=0.8{\mu}m$ wavelength and clockwise polarization is propagated along the x-direction with a focused transverse Gaussian profile. The spot size is $w_0=10\lambda$   and the pulse duration is $\tau=20$fs. The normalized laser amplitude is $a_0=eE_0/m_e{\omega}c=316/{\sqrt2}$,  corresponding to a laser electric field $E_0=8.0\times10^{14}$ V/m, as adopted in Ref. \cite{shen2007bubble}, where $e$,$m_e$ are the electron mass and charge, respectively. The simulation mesh resolution is $dx=0.03125\lambda$  and $dy=dz=0.5\lambda$, in longitudinal and transverse directions respectively. There are 4 macro-particles per species per cell. The moving window simulation box is $50\lambda\times90\lambda\times90\lambda$ with open boundary conditions in each direction.	The proton polarization is defined by its spin vector $\mathbf{s}$, which has an  absolute value of 1 and a direction calculated from the  T-BMT equation $d\mathbf{s}/dt=-\mathbf{\Omega}\times\mathbf{s}$ \cite{mane2005spin,thomas1926motion},	
	\begin{equation}\label{BTMT}
	\mathbf{\Omega}=\frac{q}{mc}[ (a_p+\frac{1}{\gamma})\mathbf{B}-(\frac{a_p\gamma}{\gamma+1})(\frac{\mathbf{v}}{c}\cdot\mathbf{B)}\frac{\mathbf{v}}{c}-(a_p+\frac{1}{1+\gamma}
	)\frac{\mathbf{v}}{c}\times\mathbf{E}] 
	\end{equation} 
	where $m,q$ are the proton mass and charge, respectively;  $a_p=1.87$ is the dimensionless anomalous magnetic moment of the proton,  $\gamma$ the Lorentz factor of the proton velocity, $c$ the light speed in vacuum, $\mathbf{B}$ the magnetic field and $\mathbf{E}$  the electric field in the laboratory frame. Equation~[\ref{BTMT}] can be evaluated using an adapted version of the standard Boris operator splitting method commonly used for the momentum integration in PIC codes\cite{hockney1988computer,birdsall2004plasma}. Although the particle spin can in principle be altered by the Stern-Gerlach and Sokolov-Ternov effects, these have been neglected in our study based on prior work of Ref.~\cite{PhysRevAccelBeams.23.064401}. The simulation geometry is shown in Figure 1, where the initial electron, proton and tritium densities are $1.5\times10^{21} cm^{-3}$, $1.0\times10^{20}cm^{-3}$ and $1.4\times10^{21}cm^{-3}$ respectively. The plasma density is uniform and the vacuum length at left edge is $10\lambda$. The proton:tritium density ratio is 1:14. The target is initially pre-polarized in the x-direction \cite{wen2019polarized}, as depicted by the red arrows in Fig. 1.

    \section{Result and discussion}		
	 An intense laser pulse injected into an underdense plasma can result in a blowout state in which the electrons are expelled by laser while the ions remain immobile. Owing to the charge separation field, the electrons return back to the axis forming a positively charged ``bubble'', which has a phase velocity equal to the group velocity of the driver laser. Increasing the plasma density to near critical density results in a lower bubble velocity and an increase of the electromagnetic field in the bubble. Because of its longer exposure to a more intense accelerating field, the motion of protons cannot be ignored,  and some protons may be  captured by the wakefield. 
	
		\begin{figure}[ht]
			\includegraphics[width=0.5\textwidth]{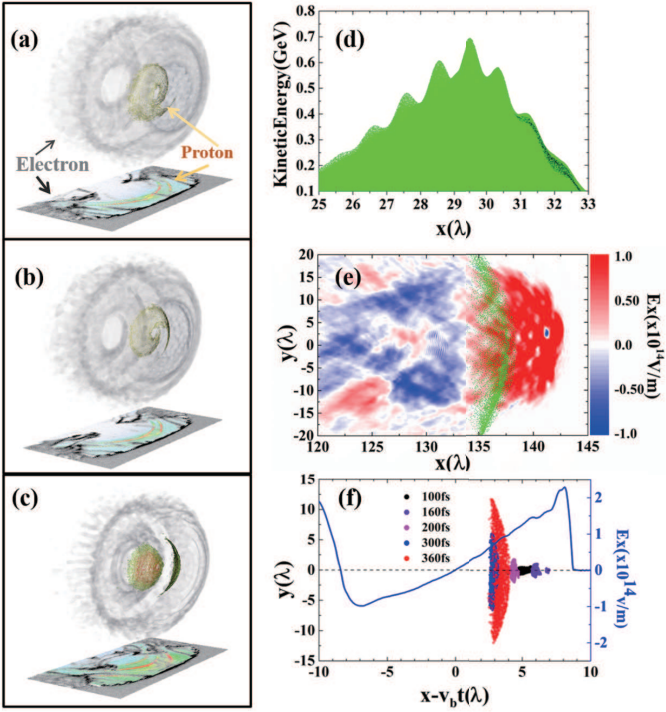}
			\centering
			\caption{
				The distribution of proton and electron density  at $100$fs in cases with different laser polarization:(a) clockwise polarization, (b) anti-clockwise polarization and (c) linear polarization. (d) The distribution of proton kinetic energy $E_k>0.1$ GeV with x at 100fs.  (e) The distribution of $E_x$ at 400fs. The protons whose kinetic energy larger than 4 GeV were denoted as green dots. (f) The position of selected protons (denoted as black dots in Fig. 2(d)) at different times in bubble frame, where $v_b$ is the bubble velocity. The distribution of the longitudinal field on the axis at 120fs is shown as a blue line. The dashed black line presented the longitudinal field $E_x=0$. The laser intensity is $a_0=316/\sqrt{2}$ and the initial electron density is $n_{e0}=1.5\times10^{21}cm^{-3}$. The proton and tritium densities are  $n_{p0}=1.0\times10^{20}cm^{-3}$ and $n_{T0}=1.4\times10^{21}cm^{-3}$, respectively. }
			\label{Figure2}
		\end{figure} 
	
	As shown in Fig. 2(a), the electron bubble is formed at $100$fs and some electrons are injected into the tail of bubble. There are also some protons distributed near the driving laser, which located at the head of bubble. The proton density forms a \textit{vortex microstructure}. The rotation direction is same as the laser polarization. For the laser amplitude $a_0=316/{\sqrt{2}}$, the quiver velocity of protons would be $v_p/c=a_0/1836=0.172$, which means that the protons can move in the laser field. During the first stage of pre-acceleration, the proton motion is therefore directly affected by the ultra-intense laser field. This can be demonstrated by examining the dependence of the resultant kinetic energy distribution on the laser polarization in the first stage. Figs. 2(b) and 2(c) show the distribution of proton density at same time with anti-clockwise and linear polarization, respectively. Compared with the case of clockwise polarization, the vortex structure follows the laser polarization such that the microstructure is imprinted directly by the laser. When the laser is linearly polarized, the proton density takes on a  multilayer instead. These features are also reflected in the proton kinetic energy ($E_k$) phase space --  Fig. 2(d) for the clockwise polarized laser. Here, the proton energy has a periodic variation with the laser wavelength, another strong indication that the protons are directly modulated by the laser.

      At later times, the protons with kinetic energy $E_k>4$ GeV are still located at regime with $E_x>0$ , as shown in Fig. 2(e). It means that the high energy protons can catch up with the electron ``bubble'' and continue to gain energy in the wakefield. To track the acceleration dynamics further, nearly 2000 protons are selected with $E_k>4$ GeV at 400fs. Their positions at $100$fs are shown as black dots in Fig. 2(d); their subsequent positions  relative to the longitudinal electric field profile at later times are displayed in Fig. 2(f). At 200fs, these protons (purple dots) are still located within the accelerating region of the wakefield with $E_x>0$, although they undergo some slippage before regaining some ground between 200fs and 400fs. In other words, the protons are trapped in the wakefield at nearly 400fs, and are accelerated continuously. On the other hand, their transverse position increases  with time, implying that the bunch experiences a defocusing radial  field. 
      
	  A further series of simulations show that the protons always slip out from the bubble regime and cannot be further accelerated if a laser with linear polarization is used. Compared to the circular polarization case, the bubble size is smaller and the longitudinal filed $E_x$ is weaker in the case of linear polarization owning to the oscillating term of ponderomotive force\cite{kruer1985j}. The protons cannot obtain enough energy to catch up the acceleration field in the bubble.

	 The original motivation of this study was to analyse the time evolution of the proton spin. The polarization of selected protons is investigated first. The distribution of the proton $s_x$ at $100$fs is plotted in Fig. 3(a). At this time, the $s_x$ values do not change significantly, remaining close to their initial value of $s_x=1$. At the same time, the spins exhibit a periodic structure at the laser wavelength. The corresponding proton positions and a map of $B_z$ are shown in Fig. 3(c). The protons are located in the laser field near the axis, where the $B_z$ of wakefield could be ignored. Here, the laser electric intensity is $\sim10^{14}$ V/m, the correspondence of  magnetic field nearly $10^6$ T. At 100fs, the kinetic energy of proton is not very large $\gamma\simeq1$. From Eq. [\ref{BTMT}], $\Omega\simeq\frac{q}{mc} (a_p+\frac{1}{\gamma})\mathbf{B}=2.67\times10^{13}$ Hz, where the cycle of rotation is nearly $23$fs, implying that the polarization could also be affected by the laser field directly.
	
	 \begin{figure}[H]
		\includegraphics[width=0.5\textwidth]{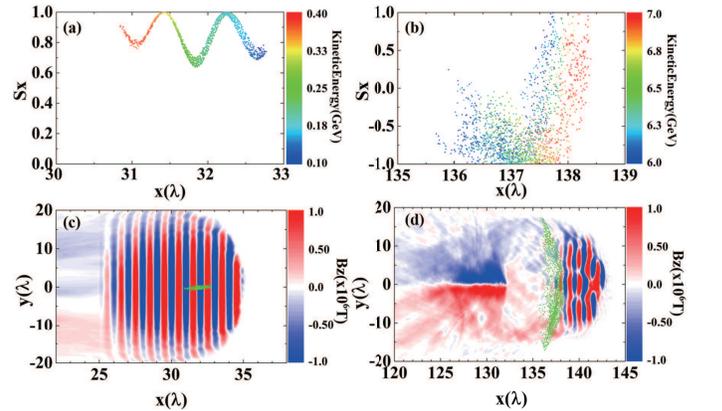}
		\centering
		\caption{ The distribution of $s_x$ with x at different time, (a)100fs and (b)400fs. The distribution of $B_x$ at the respondence time (c) 100fs and (d)400fs.}
		\label{Figure3}
	\end{figure}
			
       \begin{figure}[ht]
       	\includegraphics[width=0.4\textwidth]{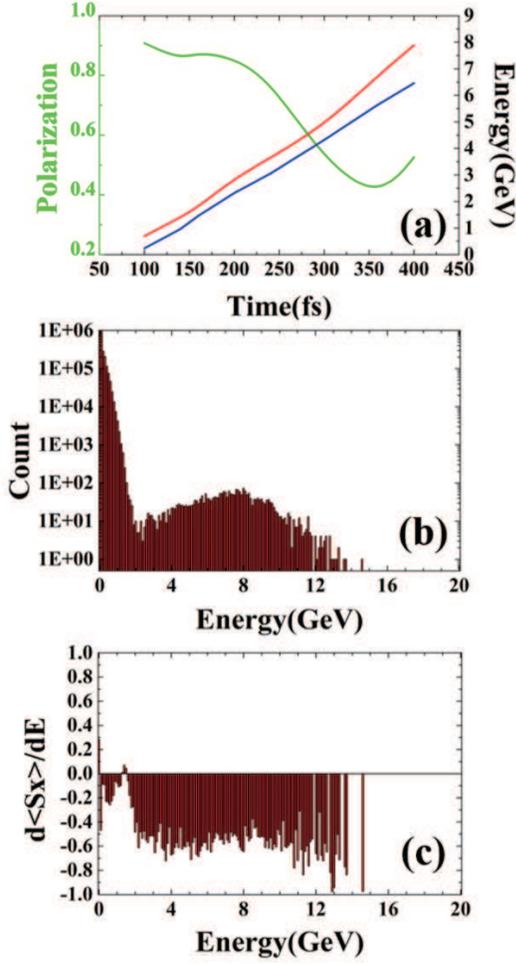}
       	\centering
       	\caption{  (a) The polarization of selected particles together with average and maximum energies. (b) The energy spectrum at 800fs and (c) the distribution of $s_x$ with energy. 
       	}
       	\label{Figure4}
       \end{figure}
 
	 By contrast, later on at $400$fs (in Fig. 3(b)), the polarization distribution of $s_x$ ranges from -1 to 1, in particular the spin goes negative for trailing protons. At this time, these protons have slipped backwards and mainly sample the field generated by the cavity wakefield. In the bubble regime, the transverse field usually focuses electrons, but  defocuses for positively charged particles, for example, positrons and protons. The transverse size of a proton bunch will therefore increase in time.  Here, it can be seen that the transverse size of protons has increased, especially its rear portion, as shown in Fig. 3(d). These are located at the sheath of the bubble, where $B_z$ is larger and the polarization rotates faster. Here, the maximum $B_z$ is nearly $0.2\times10^6$ T, so the rotation period is just 100fs for protons with $\gamma\simeq1$. For trapped protons located near the laser axis, this period will be larger, resulting in less spin rotation.

	The polarization for particle beams is defined as $P=\sqrt{\langle{s_x}\rangle^2+\langle{s_y}\rangle^2+\langle{s_z} \rangle^2}$, where $\langle{s_i}\rangle$ is the average value at each direction; for the beam comprising the previously selected particles this parameter is shown in Fig. 4(a). The average and maximum energies are also displayed here as the red and blue lines respectively. The maximum energy is nearly 8 GeV and the average energy is 6.5 GeV at 400fs. Over time,  the protons are accelerated while their polarization decreases. It should be noted that the polarization $P$ increases near $300$fs according to this definition. 
  	
  	After protons become trapped in the wakefield, they are accelerated continuously, yielding a final energy spectrum of all protons at 800fs shown in Fig. 4(b). The corresponding distribution of $\langle{s_x}\rangle$ is given in Fig. 4(c). Although the maximum energy is nearly 14.63 GeV, this is recorded by a single simulation particle, which means that $\langle{s_i}\rangle={s_i}$ and its polarization $P$ is always 1. In order to study the polarization with meaningful statistics, it is necessary to use $\langle{s_x}\rangle$ instead of $P$. As revealed in Fig. 4(c), the $\langle{s_x}\rangle$ becomes negative, which causes the protons' polarization to change rapidly via this acceleration mechanism. The number of protons with $E_k>4$ GeV was nearly $80.70$ pC. Here, $E_k>4$ GeV was used because of the proton energy cannot be higher than 4 GeV without acceleration in the wakefield. In this case, the average energy is $7.46$ GeV and the polarization is $P=0.57$.
  	
    \begin{figure}[H]
	 	\includegraphics[width=0.5\textwidth]{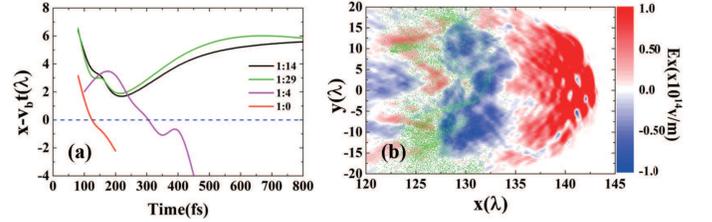}
	 	\centering
	 	\caption{ (a) The history of the distance from protons with maximum $E_k$ to position with $E_x=0$ in the cases with different ratio. (b) The distribution of $E_x$ at 400fs for the case with ratio 1:4. The protons with $E_k>3$ GeV are shown as green dots.  
		}
		\label{Figure5}
	\end{figure}
  
    The proton dynamics is sensitive to the number ratio of hydrogen to tritium, as shown in Fig. 5(a). Here, the position of $E_x=0$ is denoted as a dashed blue line. The proton positions with maximum $E_k$ for different ratios are shown by different colored lines. A position below zero means that the protons slip out from the acceleration region, and can no longer be accelerated by the wakefield. This phenomenon occurs for the proton fraction above $20\%$, a case illustrated in Fig. 5(b), where the protons with energy higher than 3 GeV at $400$ fs are located in the $E_x<0$ region of the wakefield and thus they cannot be accelerated further. 
    
   \begin{figure}[htb]
   	\includegraphics[width=0.44\textwidth]{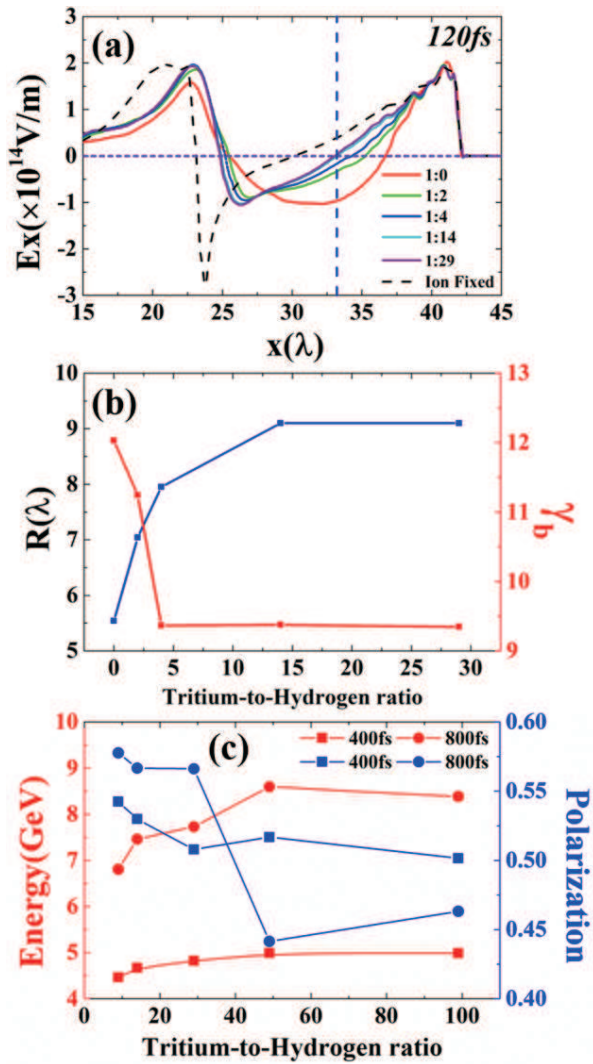}
   	\caption{(a) The distribution of $E_x$ with x for different ratios. (b) The longitudinal radius of $E_x$ and the velocity of bubble, where ${\gamma}_b=1/\sqrt{1-(v_b/c)^2}$. (c) The average energy of protons whose kinetic energy larger than 4 GeV and their polarization $P$.}
   	\label{Figure6}
   \end{figure}
	  	 
	The $E_x(x)$ profiles at 120fs are shown in Fig. 6(a) for  different ion ratios. In the case of electron acceleration in bubble regime, the ion or proton is always considered immobile, as denoted as dashed black line. For near-critical densities, the ion motion cannot be ignored. Here, we can see that the $E_x$ profile is clearly altered by the ion composition, especially its slope $\Delta{E_x}/\Delta{x}$. With increasing proportion of tritium, the distribution of $E_x$ approaches the fixed ion case. The bubble radius is defined as the distance from the head of the bubble to its center, where $E_x=0$ (dashed blue line). This parameter increases with the amount of tritium, whereas the velocity of the bubble decreases, as shown in Fig. 6(b). After being pre-accelerated directly by laser at the head of bubble, it takes longer for protons to slip back to the center of bubble with larger radius, where the proportion of protons is smaller. This  helps the proton to obtain enough energy to catch up the wakefield. Although our study has been restricted to hydrogen-tritium mixtures, our results should be equally applicable to HCl gas targets,  where the density ratio of proton and electron is 1:18. The density ratio of proton and electron is more convenient, considering the species of ion has been changed and the motion of heavier ions, either Tritium or Chlorine can be ignored.

	 Finally, we have studied that the dependence of average energy and polarization on the ion ratio, -- Fig. 6(c). Although the average $E_k$ of protons with $E_k>4$ GeV increases with lower proton fraction, their polarization decreases. This difference of polarization can be accounted for by transverse defocusing of protons in the bubble regime, which could feel more intense magnetic field. 
	
\section{Summary}
    To summarize, we have studied the generation of high-energy proton beams including polarization properties in the interaction of ultra-high-intensity lasers with near-critical density plasmas. After pre-acceleration by a circularly polarized laser, the protons are trapped in the front region of a wakefield bubble and further accelerated by the  wakefield, where the acceleration gradient is nearly $10^{14}$ V/m. The protons can gain $~10$ GeV in this field  within $100{\mu}m$. Because the laser electric field is  nearly $10^{14}$ V/m, the proton polarization can  be  affected by laser. As the transverse size of the proton bunch increases, it experiences the full bubble magnetic field up to $0.1\times10^6$ T, so its polarization is also modified by the wakefield. Finally, the relative proportion of hydrogen and tritium in the gas has a strong influence on the proton acceleration in this regime. The radius and the velocity of the accelerating field structure depend critically on the ion ratio. For sufficiently large ratios, a polarized proton beam can be trapped by the wakefield and accelerated to multi-GeV energies.

\begin{acknowledgments}		
    Simulations have been carried out on  the JURECA supercomputer at J\"{u}lich Supercomputing Centre, which is granted from the Projects JZAM04 and  Lapipe. The work of A.H. and M.B. has been carried out in the framework of the JuSPARC (Jülich Short-Pulse Particle and Radiation Center) and has been supported by the ATHENA (Accelerator Technology Helmholtz Infrastructure) consortium. This work was supported by the China and Germany Postdoctoral Exchange Program from the Office of China Postdoctoral Council and the Helmholtz Centre (Grant No: 20191016) and China Postdoctoral Science Foundation (No.
    2018M641993). This work was also supported by the Strategic Priority Research Program of Chinese Academy of Sciences (Grant No. XDA25050100), the National Natural Science Foundation of China (Grant Nos. 11975154, 11675108, 11655002, 11991074 and 11775144), the Science Challenge Project (No. TZ2018005).
\end{acknowledgments}
	
		%\bibliographystyle{unsrt} %plain
		%\bibliography{ref}	
    	%\begin{thebibliography}{99}
	
	%\end{thebibliography}

\end{document}